\documentclass[aps,pra,twocolumn,superscriptaddress]{revtex4-1}
\usepackage{graphicx}
\usepackage{amssymb}
\usepackage{amsmath}
\usepackage{bm}
\usepackage{xcolor}
 \usepackage{verbatim}

\begin{document}

\title{Defect Saturation in a Rapidly Quenched Bose Gas}

\author{Junhong Goo}
\affiliation{Department of Physics and Astronomy, Seoul National University, Seoul 08826, Korea}

\author{Younghoon Lim}
\affiliation{Department of Physics and Astronomy, Seoul National University, Seoul 08826, Korea}
\affiliation{Center for Correlated Electron Systems, Institute for Basic Science, Seoul 08826, Korea}

\author{Y. Shin}
\email{yishin@snu.ac.kr}

\affiliation{Department of Physics and Astronomy, Seoul National University, Seoul 08826, Korea}
\affiliation{Center for Correlated Electron Systems, Institute for Basic Science, Seoul 08826, Korea}
\affiliation{Institute of Applied Physics, Seoul National University, Seoul 08826, Korea}


\begin{abstract}

We investigate the saturation of defect density in an atomic Bose gas rapidly cooled into a superfluid phase. The number of quantum vortices, which are spontaneously created in the quenched gas, exhibits a Poissonian distribution not only for a slow quench in the Kibble–Zurek (KZ) scaling regime but also for a fast quench in which case the mean vortex number is saturated. This shows that the saturation is not caused by destructive vortex collisions, but by the early-time coarsening in an emerging condensate, which is further supported by the observation that the condensate growth lags the quenching in the saturation regime. Our results demonstrate that the defect saturation is an effect beyond the KZ mechanism, opening a path for studying critical phase transition dynamics using the defect number distribution.

\end{abstract}

\maketitle

When a system undergoes a continuous phase transition, spatial domains of the ordered phase randomly develop owing to the causal independence of far distant regions and topological defects are possibly formed at the interfaces of the domains~\cite{delCampo14}. The Kibble–Zurek mechanism (KZM) provides a universal framework for the above defect formation, predicting the power-law dependence of the defect density on the quench rate, in which the scaling exponent is determined by the critical exponents of the phase transition and the spatial dimensionalities of the system and defects~\cite{Kibble76,Zurek85,Hohenberg77,Kibble80,Zurek96,delCampo13,Dziarmaga10}. The KZM has been experimentally tested using many controllable systems, such as superfluid helium~\cite{Hendry94,Bauerle96,Ruutu96,Dodd98}, liquid crystals~\cite{Chuang91}, superconductors~\cite{Carmi00,Monaco02}, trapped ions~\cite{Pyka13, Ulm13, Ejtemaee13}, and ultracold atomic gases~\cite{Weiler08, Lamporesi13, Navon15,Sadler06,Ko19,Donadello16}. However, it does not provide a quantitative estimation of the defect density, which requires a full understanding of the complex phase transition dynamics, including domain emerging and coarsening after passing through the critical region~\cite{Biroli10,Das12}.

Recently, a noticeable deviation from KZM predictions was exhibited in atomic gas experiments, in which a trapped sample was quenched to undergo a normal-to-superfluid phase transition, generating quantum vortices, which became saturated for rapid quenches~\cite{Ko19,Donadello16}. A plausible explanation of the defect saturation is the fast relaxation of the defects via pair annihilation at high densities~\cite{delCampo10,Ko19,Donadello16,Liu18}. From this perspective, the KZM still holds in a fast quench regime, and the observed saturation is attributed to the experimental incapability to detect individual vortices in such a turbulent condensate. Meanwhile, a recent theoretical study anticipated that the early-time coarsening before a well-defined condensate forms is highly critical that when the quench time is shorter than the characteristic condensate growth time, the defect formation dynamics qualitatively changes, making the defect density independent of the quench time~\cite{Chesler15}. This scenario is beyond the KZM scope, highlighting the significance of the order parameter growth and the coarsening in the defect formation process.

\begin{figure}[b]
	\includegraphics[width=8.5cm]{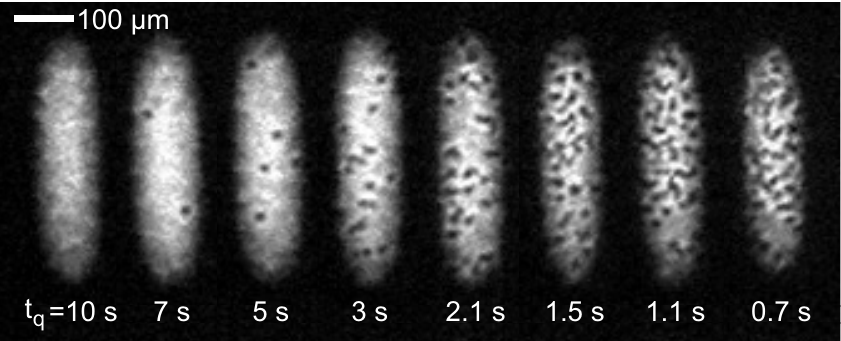}
		\caption{Spontaneous defect formation in an atomic Bose gas quenched into a superfluid phase. A sample is cooled by linearly decreasing the trap depth for variable quench time $t_\mathrm{q}$. Images of the gas for various $t_\mathrm{q}$, obtained after a time-of-flight, where quantum vortices generated during the phase transition are detected by their expanded, density-depleted cores.}
\end{figure}

\begin{figure*}[t]
	\includegraphics[width=17.0cm]{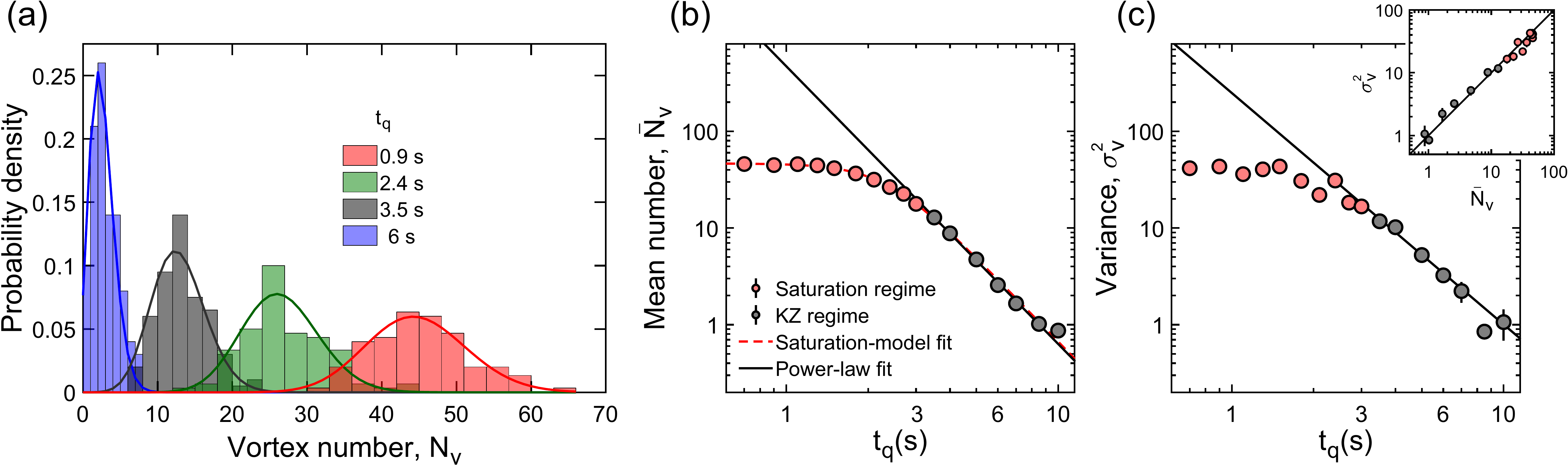}
	\caption{Vortex number statistics. (a) Probability distribution of vortex number for various quench times $t_\mathrm{q}$. Each distribution is obtained from a collection of 100 measurements for a given $t_\mathrm{q}$. The solid lines denote Poissonian curves fit to the data. (b) Mean value $\bar{N}_\mathrm{v}$ and (c) variance $\sigma^2_\mathrm{v}$ of the vortex number as functions of $t_\mathrm{q}$ on log--log axes. The dashed line in (b) is a model curve fit to the data, where saturation time $t_\textrm{sat}=2.3~\mathrm{s}$ (see text for details). The solid lines denote a power-law function fit to the data in the KZ regime of $t_\mathrm{q}\geq1.5 t_{\mathrm{sat}}$ (grey circles). Inset in (c) displays the measurement results in plane of $\bar{N}_\mathrm{v}$ and $\sigma^2_\mathrm{v}$, and the solid line represents a linear fit to them, yielding $\sigma^2_\mathrm{v}=1.01(12)\bar{N}_\mathrm{v}$. The error bars in (b) and (c) indicate the standard errors. If the error bars are invisible, they are shorter than the marker size.}
\end{figure*}

In this paper, we experimentally investigate the defect saturation occurring in a rapidly quenched $^{87}$Rb Bose gas by measuring the vortex number distribution and the condensate growth curve for various quench times. The vortex number exhibits a Poissonian distribution in the entire range of the quench times, from the Kibble–Zurek (KZ) scaling regime to the saturation regime. This study excludes the defect saturation scenario based on destructive vortex collisions. Concurrently, in the saturation regime, the condensate growth is significantly delayed after the quench, consistent with the early-time coarsening scenario~\cite{Chesler15}. The scaling of the saturated defect density with the vortex core area is found to be comparable to previous values measured for strongly interacting Fermi gases~\cite{Ko19}, suggesting an interesting possibility of a universal defect formation dynamics for the rapid quench regime. This study provides experimental evidence for a beyond-KZM effect and opens a new path for studying critical phase transition dynamics using the defect number distribution~\cite{Gomez20}.

We start our experiment by preparing a cold thermal cloud of $^{87}$Rb in an optical dipole trap (ODT) having a highly oblate and elongated geometry. The initial sample has $4.1(1)\times10^{7}$ atoms, and its temperature is $440(18)$~nK. The sample is quenched via evaporative cooling by linearly lowering the trap depth from $U_{\mathrm{i}}=1.15U_{\mathrm{c}}$ to $U_{\mathrm{f}}=0.33U_{\mathrm{c}}$ for a variable quench time $t_{\mathrm{q}}$. Here, $U_{\mathrm{c}}$ is the critical trap depth for Bose–Einstein condensation in an equilibrium sample, where the atom number is decreased to $\approx 3.7\times10^{7}$. The quench time $t_{\mathrm{q}}$, varies from 0.7 to 10~s. In the explored range of $t_\mathrm{q}$, the linear relationship between the sample temperature and the trap depth is examined to ensure that the quench rate is quantified well by the inverse of $t_\mathrm{q}$~\cite{SM}. The collision time of atoms is $\approx 1.6$~ms for the peak atom density of the sample at the critical point. At the end of the quench, the atom number is typically $\approx 1.1\times10^{7}$ and the sample temperature is $\approx 50$~nK.

As the sample is cooled to undergo a phase transition, superfluid phase domains with independent broken-symmetry develop therein. In the highly oblate sample, owing to fast thermalization along the tight axial confining direction, the phase domains are mainly formed in the radial plane, and their merging leads to generation of quantum vortices, whose rotation axes are aligned parallel to the axial direction~\cite{Ko19, footnote1}. To facilitate the domain formation and merging dynamics, a hold time $t_{\mathrm{h}}=1.25~\mathrm{s}$ is applied after the quench~\cite{SM}. The created vortices are detected by imaging the sample after releasing the trapping potential and letting their density-depleted cores expand during a time-of-flight of 40.9~ms (Fig.~1).

Our experiment uses a large-area sample, which allows generation up to several tens of vortices for rapid quenches. This number is an order of magnitude higher than those in previous KZ experiments with weakly interacting Bose gases~\cite{Weiler08,Lamporesi13,Donadello16} and comparable to that in a recent experiment on strongly interacting Fermi gases~\cite{Ko19}. The ODT is constructed by focusing a truncated Gaussian laser beam, where the trapping potential is elongated and flattened along the beam axis direction by beam clipping~\cite{Lim21,SM}. The Thomas--Fermi radii of the fully grown condensate are $R_{x,y,z}\approx (63,235,2.6)~\mu\mathrm{m}$. Such a large-area sample is exceedingly beneficial for obtaining reliable vortex number statistics.

The representative histograms of the vortex number, $N_\textrm{v}$, for various quench times are presented in Fig.~2(a). $N_\textrm{v}$ is counted manually and each histogram is obtained from an ensemble of 100 image data. As $t_\mathrm{q}$ decreases, the mean vortex number increases and accordingly, the width of the number distribution also increases. We find that the measurement results are described remarkably well by a Poissonian distribution for each $t_\mathrm{q}$. This demonstrates the spatial randomness of the spontaneous defect formation process in the quenched system. Poissonian distributions of defect numbers were marginally observed in a previous experiment with elongated samples~\cite{Donadello16} and numerically predicted for a two-dimensional (2D) system~\cite{delCampo21}.

Figure~2(b) displays the mean vortex number, $\bar{N}_{\mathrm{v}}$, as a function of the quench time, $t_{\mathrm{q}}$. For long $t_{\mathrm{q}}$, $\bar{N}_{\mathrm{v}}$ shows the characteristic power-law dependence on $t_{\mathrm{q}}$ as predicted by the KZM. However, when $t_\mathrm{q}$ decreases below approximately $3~\textrm{s}$, it becomes saturated, as observed in previous experiments~\cite{Ko19,Donadello16}. To systematically distinguish between the two regimes, we adopt a phenomenological model of the vortex number $N_{\mathrm{S}}(t_\mathrm{q}) =N_{\mathrm{sat}}[1+(t_{\mathrm{q}}/t_{\mathrm{sat}})^{\delta\beta_{\mathrm{KZ}}}]^{-1/\delta}$, where $N_{\mathrm{sat}}$ denotes the saturated defect number, $t_{\mathrm{sat}}$ the saturation quench time, $\beta_{\mathrm{KZ}}$ the scaling exponent, and $\delta$ a heuristic parameter for tuning the transition from the scaling behavior to saturation. A model curve fit to the experimental data yields $N_{\mathrm{sat}}=46.4(1)$, $t_{\mathrm{sat}}=2.3(3)$, and $\beta_{\mathrm{KZ}}=2.9(5)$ with $\delta=1.4(5)$. In the following, we refer to the regions with $t_{\mathrm{q}}\geq1.5t_{\mathrm{sat}}$ and $t_{\mathrm{q}}<1.5t_{\mathrm{sat}}$ as KZ scaling and saturation regimes, respectively. A power-law function of $\bar{N}_{\mathrm{v}}= C {t_{\mathrm{q}}}^{-\alpha_{\mathrm{KZ}}}$ fit to the data in the KZ regime yields $\alpha_{\mathrm{KZ}}=2.9(2)$, which is equal to the value of $\beta_{\mathrm{KZ}}$ within uncertainty. 

In Fig.~2(c), we display the variance of the vortex number, $\sigma^2_{\mathrm{v}}$, as a function of $t_{\mathrm{q}}$. Similar to $\bar{N}_{\mathrm{v}}$, $\sigma^2_{\mathrm{v}}$ shows power-law and saturation behavior in the KZ and saturation regimes, respectively. The power-law exponent in the KZ regime is measured to be $\alpha_{\sigma^2}=2.4(6)$, similar to $\alpha_{\mathrm{KZ}}$ within fit error. The inset in Fig.~2(c) displays our experimental data in the plane of $\bar{N}_{\mathrm{v}}$ and $\sigma_\mathrm{v}^2$, and a linear fit yields $\sigma^2_{\mathrm{v}}=1.01(12)\bar{N}_{\mathrm{v}}$, corroborating the Poissonian distribution of the vortex number.

For point defect generation in a homogeneous 2D system, the KZM predicts the power-law exponent to be $\alpha_\mathrm{KZ}=2\nu/(1+\nu z)$, where $\nu$ and $z$ are the static and dynamic critical exponents of the phase transition, respectively~\cite{Zurek85,Hohenberg77}, whose values recently measured with weakly interacting Bose gases~\cite{Navon15,Donner07} suggest $\alpha_\mathrm{KZ}\approx 2/3$. Our measured value of $\alpha_{\mathrm{KZ}}$ is much higher than the prediction, which might be attributed to the density inhomogeneity of the trapped sample~\cite{delCampo11}. Del Campo {\it et al.}~\cite{delCampo11} presented a theoretical estimation of the KZ exponent for an inhomogeneous system trapped in a power-law potential. Assuming that the ODT potential is quadratic and quartic along the short and long axes, respectively~\cite{Lim21}, the exponent is estimated to be $\approx 16/9$~\cite{SM}. In a harmonic trap case, which is conceivable because of the small non-zero curvature of the trapping potential at its center, the exponent is given to be $\approx 7/3$, which is still small to account for the measurement result. The power-law trap assumption might be inapplicable to the non-Gaussian ODT regarding the KZM. Quantitative understanding of the measured value of $\alpha_{\mathrm{KZ}}$ requires further experimental and theoretical investigations.

\begin{figure}[t]
	\includegraphics[width=8.2cm]{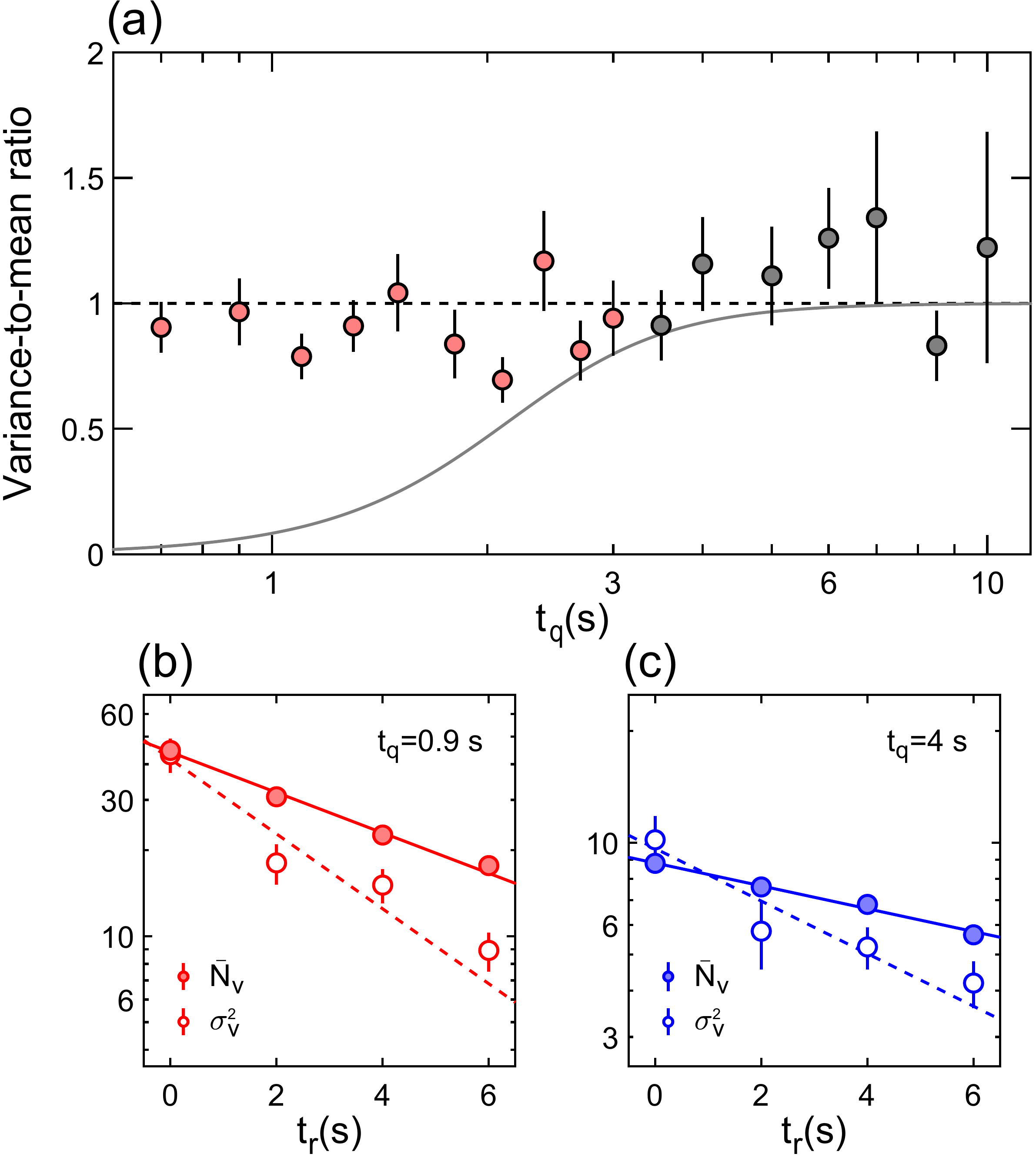}
	\caption{(a) Variance-to-mean ratio (VMR) of the vortex number as a function of $t_\mathrm{q}$. The solid line represents a model curve based on the vortex decay scenario, where the initial mean vortex number is given by the KZ scaling prediction (see text for details). (b)(c) Relaxation of the vortex number distribution. $\bar{N}_\mathrm{v}$ (solid) and $\sigma^2_\mathrm{v}$ (open) are displayed as functions of additional holding time $t_\mathrm{r}$ on log–linear axes for (b) $t_\mathrm{q}=0.9~\mathrm{s}$ and (b) $4~\mathrm{s}$. Each data point was obtained from 100 realizations of same experiment. The error bars in (a)–(c) are the standard errors of the measured quantities.}
\end{figure}

Regarding the origin of defect saturation for rapid quenches, a Poissonian distribution of the vortex number in the saturation regime has an important implication. In the vortex relaxation scenario, it is assumed that the vortices are initially generated with a hypothetical mean number, ${N}_{\mathrm{KZ}}$, provided by the KZ scaling prediction and decay to ${N}_{\mathrm{S}}=\gamma N_{\mathrm{KZ}}$ with a reduction factor of $\gamma$. In this case, the vortex number variance will evolve from $\sigma^2_{\mathrm{KZ}}=N_{\mathrm{KZ}}$ to $\sigma_\mathrm{v}^2=\gamma^2\sigma^2_{\mathrm{KZ}}=\gamma {N}_{\mathrm{S}}$ because the variance decays twice as fast as the mean value. Consequently the vortex number distribution will lose its Poissonian character over time with the decrease in the variance-to-mean ratio (VMR) from unity to $\gamma<1$. In Fig.~3(a), the measured VMR is plotted as a function of $t_{\mathrm{q}}$ and compared to the model curve of $\gamma(t_{\mathrm{q}})={N}_{\mathrm{S}}/N_{\mathrm{KZ}}$ from the vortex decay scenario with ${N}_{\mathrm{KZ}}(t_\mathrm{q})= N_{\mathrm{sat}}({t_{\mathrm{q}}/t_{\mathrm{sat}}})^{-\beta_{\mathrm{KZ}}}$. The VMR tends to be close to unity, clearly excluding the effect of destructive vortex collisions in the saturation.

In Figs.~3(b) and (c), the relaxation curves of $\bar{N}_{\mathrm{v}}$ and $\sigma^2_{\mathrm{v}}$ as functions of an additional hold time $t_{\mathrm{r}}$ are displayed for $t_\mathrm{q}=0.9$~s and 4~s, respectively. Applying exponential curve fits, the decay rates of $\bar{N}_\mathrm{v}$ and $\sigma^2_\mathrm{v}$ are measured to be approximately 0.16~$\textrm{s}^{-1}$ and 0.30~$\textrm{s}^{-1}$ (0.07~$\textrm{s}^{-1}$ and 0.16~$\textrm{s}^{-1}$), respectively, for the fast (slow) quench, and as expected, $\sigma^2_{\mathrm{v}}$ decays almost twice as fast as $\bar{N}_{\mathrm{v}}$. Note that the decay rates are higher for larger $\bar{N}_\mathrm{v}$, which provides the experiential basis of the vortex decay scenario for the defect saturation. The slight decreasing tendency of the measured VMR for short $t_\mathrm{q}$ could be attributed to the relaxation effect in the hold time after the quench.

Because the vortex decay scenario is eliminated, the early-time coarsening dynamics should be responsible for the observed saturation effect. When a system is quenched through the critical point, the system is effectively frozen in terms of its spatial correlations owing to the divergence of the relaxation time of the system near the critical point, and its evolution is resumed at a time $t_{\mathrm{freeze}}\propto\sqrt{t_{\mathrm{q}}}$ after passing the critical point~\cite{Navon15,freeze}. Defects are not immediately generated at this unfreeze time because of the absence of an order parameter. A finite time is required for the order parameter to emerge, during which its spatial correlations are simultaneously coarsened, resulting in a reduction in the defect density~\cite{Biroli10,Das12}. 

Although the defect density is suppressed by the early-time coarsening, it has been generally regarded that its power-law scaling holds because the characteristic order parameter growth time, $t_{\mathrm{g}}$, scales with $t_{\mathrm{freeze}}$. However, a recent theoretical study predicted that the coarsening effect is much larger than expected; therefore, it will limit the defect density for sufficiently fast quenches~\cite{Chesler15}. More explicitly, when a quench is finished much earlier than a condensate emerges, the evolution of the condensate effectively occurs at a constant temperature; therefore, $t_\mathrm{g}$ becomes insensitive to the quench time, leading to defect saturation. The correlation between the defect saturation and the scale breaking of $t_\mathrm{g}$ was also numerically suggested in~\cite{Sonner15,Liu18}.

\begin{figure}[t]
	\includegraphics[width=8.4cm]{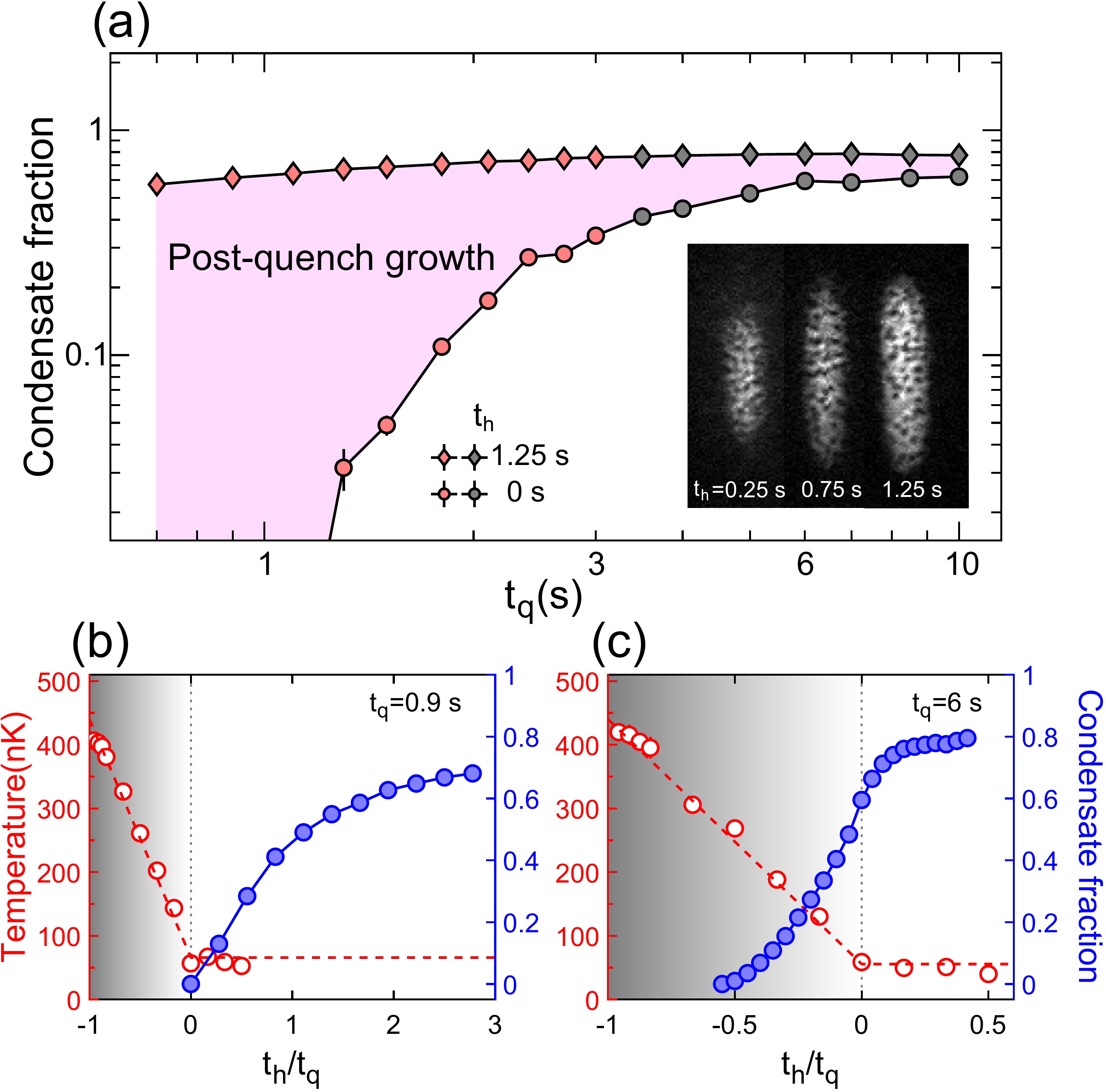}
	\caption{Post-quench condensate growth. (a) Condensate fraction at the end of the quench (circle, $t_\mathrm{h}=0$) and after a hold time of $t_\mathrm{h}=1.25~\textrm{s}$ (diamond) as functions of $t_\mathrm{q}$ on log-log axes. Inset shows images of condensates at various hold times $t_\mathrm{h}$ after a quench with $t_\mathrm{q}=1.3$~s. (b)(c) Time evolution of the temperature (open circle) and the condensate fraction (solid circle) of the sample in the quenching for (b) $t_\mathrm{q}=0.9~\mathrm{s}$ and (c) $6~\mathrm{s}$. $t_\mathrm{h}/t_\mathrm{q}=-1 (0)$ corresponds to the start (end) of the quench. The dashed lines are guides to the eyes for the temperature evolution. Each data point in (a)–(c) was obtained from ten realizations of same experiment, except those marked by diamonds in (a), which were from 100 measurements. Error bars indicate standard errors of the mean. If the error bars are invisible, they are shorter than the marker size.}
\end{figure}

To characterize the post-quench condensate growth, in Fig.~4(a), the condensate fraction measured at the end of the quench ($t_\mathrm{h}=0$) is plotted as a function of $t_\mathrm{q}$ with that at $t_\mathrm{h}=1.25~\textrm{s}$ when we count the vortex number. Clearly, the post-quench condensate growth prevails in the saturation regime. For $t_\mathrm{q}< t_\mathrm{c}\approx 1.3~\textrm{s}$, the condensate fraction at $t_\mathrm{h}=0$ is even below our detection limit of a few percent. Figs.~4(b) and (c) present the comparison of the evolution curves of the temperature and the condensate fraction for a fast quench with $t_\mathrm{q}=0.9~\textrm{s}< t_\mathrm{c}$ and a slow quench with $t_\mathrm{q}= 6~\textrm{s}$. Here, the data points for negative $t_\mathrm{h}$ are measured by abruptly interrupting the quench. In the deep saturation regime with $t_\mathrm{q}<t_\mathrm{c}$, as predicted, the temperature is constant during the condensate growth [Fig.~4(b)], so the coarsening dynamics should be qualitatively different from that in a typical quench in which the temperature continually decreasing [Fig.~4(c)].

Finally, we discuss the saturation value of the defect density. In recent experiments with strongly interacting Fermi gases in a harmonic trap~\cite{Ko19}, the defect density, $n_\mathrm{v}$, was observed to be saturated at $n_\mathrm{v} \xi_{\mathrm{v}}^2\approx 0.5 \times 10^{-4}$, where $\xi_{\mathrm{v}}=\frac{\hbar}{m_\mathrm{p} v_\mathrm{c}}$ is the effective vortex core size, with $m_\mathrm{p}$ being the fermion pair mass and $v_\mathrm{c}$ the Landau critical velocity. In our experiment with weakly interacting Bose gases, $v_\mathrm{c}$ is given by the speed of sound $c_\mathrm{s}=\sqrt{\frac{2\mu}{3m}}$ in the highly oblate geometry, with $\mu$ being the condensate chemical potential and $m$ the atomic mass~\cite{Stringari98,Kim20}, and $\xi_{\mathrm{v}}=\frac{\hbar}{m c_\mathrm{s}}\approx0.47~\mu \mathrm{m}$. For the saturated density $n_{\mathrm{v}}=\frac{N_{\mathrm{sat}}}{\pi R_x R_y}$, we obtain $n_{\mathrm{v}} \xi_{\mathrm{v}}^2\approx2.2 \times 10^{-4}$, which is approximately four times higher than the previous value. We note that in the previous experiment, the saturation value of $n_\mathrm{v} \xi_{\mathrm{v}}^2$ is slightly increased as the system evolves from unitarity into the Bose-Einstein condensate regime. It was anticipated that the saturated defect density would be governed by the final temperature of the system~\cite{Chesler15}; however, this is difficult to test in the current setting using an inhomogeneous sample, where the sample area is affected by the temperature.

In summary, we have investigated defect saturation in a rapidly quenched atomic Bose gas. The vortex number distribution was observed to be Poissonian in the saturation regime, which eliminated the vortex relaxation scenario. This consequently indicated that the defect saturation is a beyond-KZM effect, signifying the critical role of the early-time coarsening in the spontaneous defect formation dynamics. An extension of this study will be to quantitatively investigate the saturated defect density and the underlying coarsening dynamics by performing a similar set of measurements with homogeneous samples, where the atomic density of the sample and the temperature trajectory of the quench will be controlled without affecting the sample area.

\begin{acknowledgments}
We thank  J.W. Park, B. Ko, and A. del Campo for discussion, and Y. Lee for his assistance in the vortex number counting. This work was supported by the National Research Foundation of Korea (NRF-2018R1A2B3003373, NRF-2019M3E4A1080400) and the Institute for Basic Science in Korea (IBS-R009-D1). 
\end{acknowledgments}

\clearpage
\newpage

\section*{Supplemental Material}

\setcounter{equation}{0}
\setcounter{figure}{0}
\setcounter{table}{0}
\makeatletter
\renewcommand{\theequation}{S\arabic{equation}}
\renewcommand{\thefigure}{S\arabic{figure}}
\renewcommand*{\bibnumfmt}[1]{[S#1]}
\renewcommand{\thesubsection}{A\arabic{subsection}}

\subsection*{Preparation of large-area sample}

To make a sample with large area, we use a clipped Guassian optical dipole trap (ODT), described in Ref.~\cite{Lim21S}. A 1064-nm Gaussian laser beam is symmetrically truncated by a horizontal slit and vertically focused through a cylindrical lens to form a highly oblate ODT. The $1/e^2$ diameter of the beam is $\approx 22$~mm, the slit width is $\approx 14$~mm, and the focal length of the cylindrical lens is 100~mm. For the laser beam clipping, the focal region is elongated along the beam axis. The trapping potential of the ODT is characterized in a form of 
\begin{equation}
V(x,y,z)=\frac{1}{2}m[\omega_{x}^2 {x^2}+\omega_{z0}^2(1+B y)^2 z^2]+A|{y}|^{3.9}, 
\label{eq1}
\end{equation}
where the $y$ axis represents the beam propagation direction, and $z$ the vertical, gravitational direction. The trapping frequency along the $z$ axis has weak dependence on the axial $y$ position with $B=5.4(3)\times 10^{-4}~\mu\textrm{m}^{-1}$, and the axial trapping potential is approximately quartic. At the final ODT condition in the quench with the trap depth of $U=U_\mathrm{f}$, $\omega_x=2\pi\times 6.8(1)$~Hz and $\omega_{z0}=2\pi\times 164(1)$~Hz. For a quench with $t_\textrm{q}=6$~s, the fully grown condensate contains approximately $8.9\times 10^6$ atoms, and its Thomas-Fermi radii are measured to be $R_{x,y}\approx (63,235)~\mu\mathrm{m}$. The trap parameter $A$ is provided by $A=\mu/R_y^{3.9}$, where $\mu=\frac{1}{2}m\omega_x^2 R_x^2\approx k_\textrm{B}\times 38~\textrm{nK}$ ($k_\mathrm{B}$ is Boltzmann constant) is the condensate chemical potential. For the condensate in the highly oblate geometry, the speed of sound is estimated to be $c_s=\sqrt{\frac{2\mu}{3m}}\approx1.6~\textrm{mm/s}$~\cite{Stringari98S,Kim20S}.

\begin{figure}[b]
	\includegraphics[width=6.0cm]{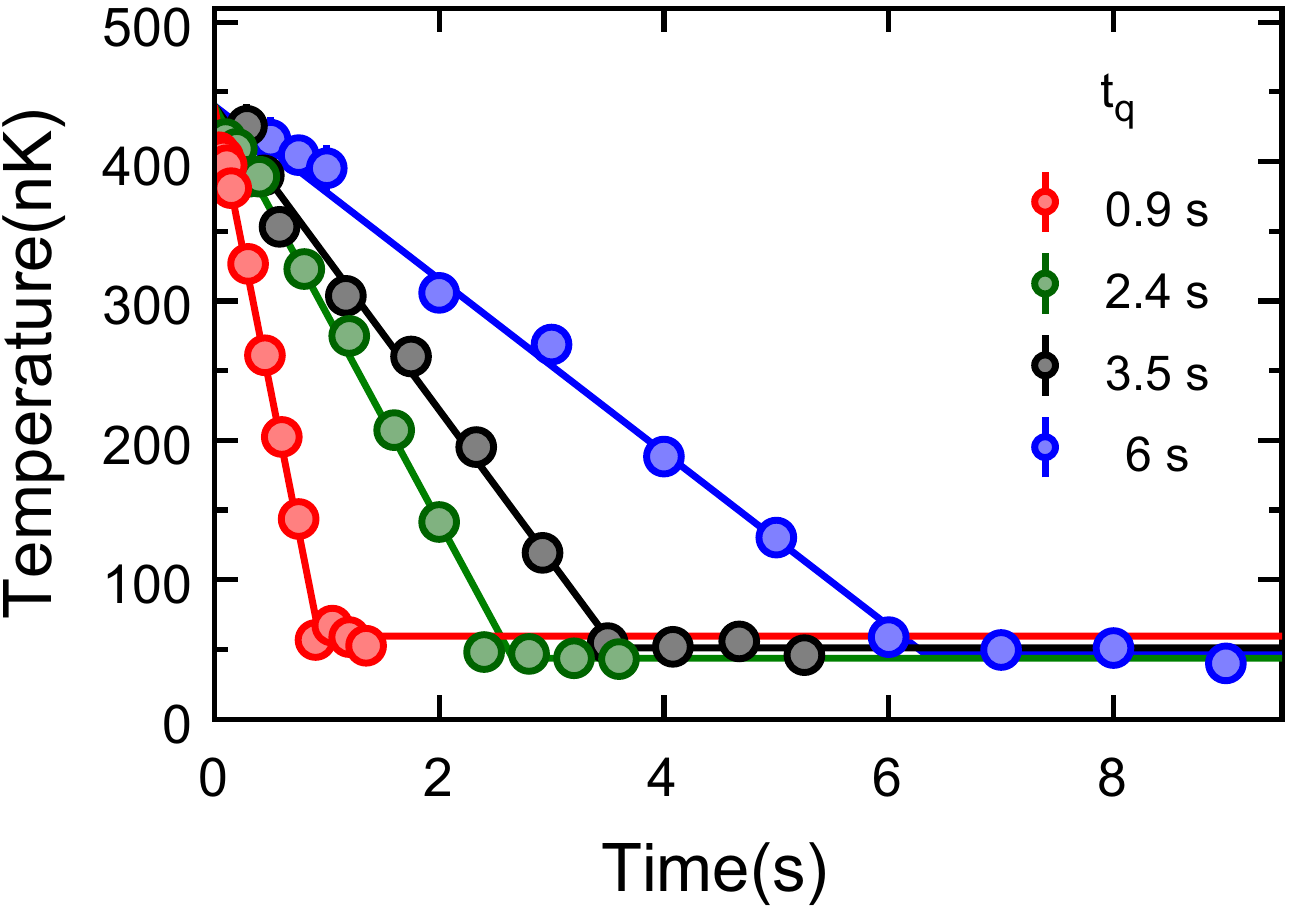}
	\caption{Evolution of the sample temperature in the quenching for various quench times $t_\mathrm{q}$. The solid lines are bilinear fits to the data for each $t_\mathrm{q}$, where the later line segments are kept horizontal. Each data point was obtained from ten measurements of same experiment, and its error bar represents the standard error of their mean.
}
\end{figure}

\subsection*{Temperature evolution in quenching}

In Fig.~S1, we display the evolution of the sample temperature, $T$, in the quenching for various $t_\mathrm{q}$. It is observed that the temperature linearly decreases during the quench and remains constant after the quench is completed, indicating the linear relationship between the sample temperature and the ODT beam intensity. The temperature is determined from the expansion rate of the thermal cloud after releasing the trapping potential. For a freely expanding thermal gas, the Gaussian width of the cloud evolves as $R_{t}(t)=\sqrt{R_{0}^2+\frac{k_\mathrm{B}T}{m} t^2}$, where $R_{0}$ is the initial width of the cloud in the trap and $t$ is the expansion time. We measure the widths, $R_1$ and $R_2$, at two different times, $t_1=40.9$~ms and $t_2=22.6$~ms, respectively, and determine the temperature as
\begin{equation}
T=\cfrac{m(R_{1}^2-R_{2}^2)}{k_\mathrm{B}(t_1^2-t_2^2)}.
\label{eq2}
\end{equation}
The thermal cloud width is measured from a Gaussian function fit to the outer wings of the 1D density profile of the sample, which is obtained by integrating the 2D column density distribution along the $y$ axis in absorption image.

\subsection*{Condensate growth and vortex formation}

To have vortices formed and discernible in a quenched sample, a certain amount of hold time is required after the quench. However, the hold time must be kept as short as possible to avoid the undesirable effects from the vortex relaxation after their formation; the vortex relaxation can not only reduce the vortex number but also affect the characteristics of the vortex number distribution. In Figs.~S2(a) and (b), the time evolution of the condensate fraction and the vortex number are displayed as functions of the hold time $t_\mathrm{h}$ for various $t_\mathrm{q}$. The condensate fraction is measured from a bimodal fit to the integrated 1D density distribution of the sample, which is obtained after 40.9~ms time-of-flight. For short $t_\mathrm{q}<t_\mathrm{c}\approx 1.3~\textrm{s}$, the emergence of a condensate is delayed even after the quench is finished and consequently, the vortex number rapidly increases during the holding period. In our main experiment, the vortex number measurements are performed at the hold time $t_{\mathrm{h}}=1.25~\mathrm{s}$, when the vortex number is fully increased for our shortest quench time. Note that the system does not reach equilibrium yet at the moment with the condensate fraction increasing but the vortex number is observed to decay already thereafter. In the saturation regime, the vortex decay rate is measured to be approximately 0.16~s$^{-1}$ [Fig.~3(b)].

\begin{figure}
	\includegraphics[width=6cm]{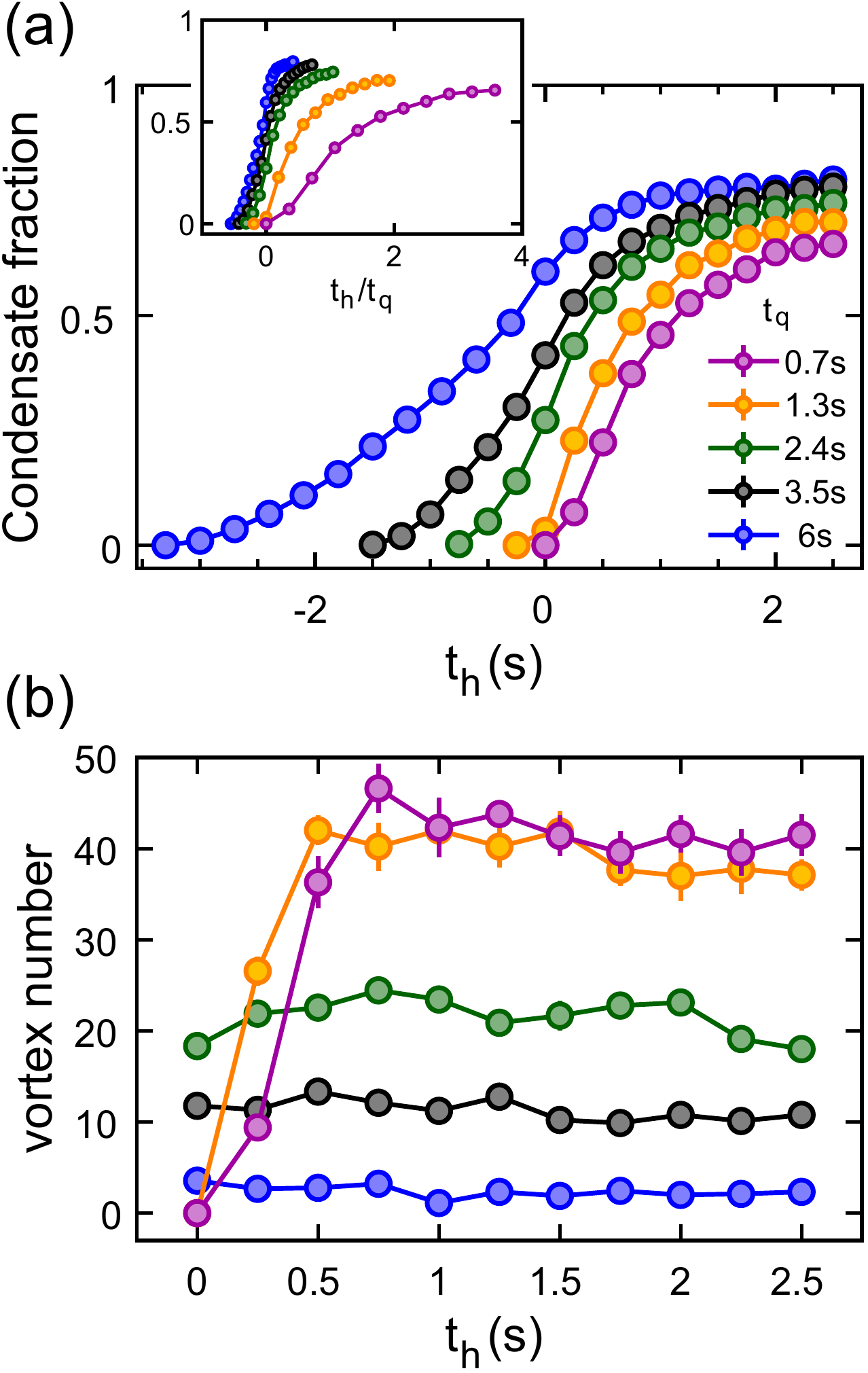}
	\caption{(a) Condensate fraction as a function of the hold time $t_\mathrm{h}$ for various $t_\mathrm{q}$. $t_\mathrm{h}<0$ denotes a time before completing the quench. The inset displays the same data as a function of $t_\mathrm{h}/t_\mathrm{q}$. (b) Evolution of the vortex number after the quench. Each data point comprises ten realizations of same experiment, and its error bar represents the standard error of the mean.
}
\end{figure}

\subsection*{KZ exponent for inhomogeneous sample}

For a trapped sample, the critical temperature has spatial dependence owing to the density inhomogeneity and in quenching, phase transition occurs at different times in different regions of the system. In the case of a power-law trapping potential of $V(x,y)=A_x |x|^{n_x} + A_y |y|^{n_y}$, the local critical temperature is given by 
\begin{equation}
T_\mathrm{c}(x,y)\sim T_\mathrm{c0} e^{-(A'_x x^{n_x}+ A'_y y^{n_y})},
\label{eq4}
\end{equation}
where $T_\mathrm{c0}$ is the critical temperature at the trap center, and for a temperature quench rate of $\sim T_\mathrm{c0}/t_\mathrm{q}$, the speed of the phase transition front along the $\alpha \in \{x,y\}$ direction is provided by 
\begin{equation}
u_{\alpha}\sim\frac{T_\mathrm{c0}}{t_\mathrm{q}}\left|\frac{\partial T_\mathrm{c}}{\partial \alpha}\right|^{-1}\sim|\alpha|^{1-n_\alpha}t_\mathrm{q}^{-1}.
\label{eq5}
\end{equation}
According to the discussion in Ref.~\cite{delCampo11S}, condensates with locally chosen phases are generated only where the speed of the phase transition front is faster than the propagation speed of the phase information, resulting in causal independence between neighboring regions. The propagation speed of the phase information is characterized by $\hat{v}=\hat{\xi}/\hat{\tau}$, where $\hat{\xi}$ is the correlation length and $\hat{\tau}$ is the relaxation time of the system. In the KZM description, $\hat{\xi}\sim t_\mathrm{q}^{\frac{\nu}{1+\nu z}}$ and $\hat{\tau}\sim t_\mathrm{q}^{\frac{\nu z}{1+ \nu z}}$ with $\nu$ and $z$ being the static and dynamic critical exponents of the phase transition, respectively, yielding $\hat{v}\sim t_\mathrm{q}^{\frac{\nu(1-z)}{1+\nu z}}$. Ignoring the spatial variations of $\hat{v}$ in the sample, the requirement of $u_\alpha> \hat{v}$ is satisfied for $|\alpha| <\alpha_* \sim t_\mathrm{q}^{\frac{1}{1-n_\alpha} \frac{1+\nu}{1+\nu z}}$, and the effective area, $S_*$, of the region where topological defects can be generated is 
\begin{equation}
S_*\sim x_*y_*\sim t_\mathrm{q}^{(\frac{1}{1-n_x}+\frac{1}{1-n_y})\frac{(1+\nu)}{1+\nu z}}.
\end{equation} 
Consequently, the expected defect number is estimated as $\bar{N}_{\mathrm{v}}\sim S_*/\hat{\xi}^2$, yielding the KZ exponent as
\begin{equation}
\alpha_{\mathrm{KZ}}=\frac{(\frac{1}{n_x-1}+\frac{1}{n_y-1})(1+\nu)+2\nu}{1+\nu z}.
\end{equation}
In the limit of $n_{x,y}\rightarrow\infty$, it approaches to the value for a homogeneous system, $\alpha_{\mathrm{KZ}}=\frac{2\nu}{1+\nu z}$. For a sample in a harmonic trap with $n_{x,y}=2$, $\alpha_{\mathrm{KZ}}=\frac{2(1+2\nu)}{1+\nu z}$~\cite{delCampo11S}. In our experiment, the trapping potential is quadratic for $x$ axis and almost quartic for $y$ axis. For $n_x=2$ and $n_y=4$, $\alpha_{\mathrm{KZ}}=\frac{2(2+5\nu)}{3(1+\nu z)}$, and using the recently measured values of $\nu \approx 2/3$~\cite{Donner07S} and $z=3/2$~\cite{Navon15S}, $\alpha_{\mathrm{KZ}}\approx16/9$.


\begin{references}
\bibitem{footnote1} For comparison, in the case of a highly prolate sample tightly confined in the radial direction, solitons are generated out of axial phase domains~\cite{Donadello16}.
\bibitem{freeze} The unfreeze time is estimated to be approximately 55~ms for our fastest quench~\cite{Navon15}, much shorter than the quench time; therefore, the extreme case of $t_\mathrm{q}<t_\mathrm{freeze}$ is not relevant in our study.
\end{references}
\end{document}